\documentstyle[12pt]{article}
\begin{document}

\begin{titlepage}
\title{Hierarchy of time scales and quasitrapping in the $N$-atom
micromaser}
\author{Georgii Miroshnichenko${}^\dag$, Andrei Rybin${}^\ddag$, Ilia Vadeiko${}^\dag$${}^\ddag$, \\
and Jussi Timonen${}^\ddag$\\$\dag$Fine Mechanics and Optics
Institute\\ Sablinskaya 14,  St. Petersburg, Russia\\
$\ddag$University of Jyv\"askyl\"a, Department of Physics\\
 PO Box 35,
Jyv\"askyl\"a, Finland}
\date{}

\maketitle

\begin{abstract}
We study the dynamics of the reduced density matrix(RDM) of the
field in the micromaser. The resonator is pumped by $N$-atomic
clusters of two-level atoms. At each given instant there is only
one cluster in the cavity. We find the conditions of the
independent evolution of the matrix elements of RDM belonging to a
(sub)diagonal of the RDM, i.e. conditions of the diagonal
invariance for the case of pumping by $N$-atomic clusters. We
analyze the spectrum of the evolution operator of the RDM and
discover the existence of the quasitrapped states of the field
mode. These states exist for a wide range of number of atoms in
the cluster as well as for a broad range of relaxation rates. We
discuss the hierarchy of dynamical processes in the micromaser and
find an important property of the field states corresponding to
the quasi-equilibrium: these states are close to either Fock
states or to a superposition of the Fock states. A possibility to
tune the distribution function of photon numbers is discussed.
\end{abstract}
\vskip 0.5cm
 PACS number 42.50

 communicating author: Andrei.Rybin@phys.jyu.fi
\end{titlepage}

Recent  developments in the cavity electrodynamics \cite{1} gave rise to the creation of a real physical device - micromaser which operates on highly excited Rydberg atoms pumped through a high-Q resonator \cite{2}.
Existing literature mostly focuses on the ideal (basic) model which is the so-called one-atom micromaser \cite{3,4,5,6}. This device is assumed to operate in such a way that no more than one atom excited with the  probability 1 can be found in the cavity at each given instant of time.
The basic model is justified by the following assumptions: the average velocity of injection $R$ ,  and the time of interaction $\tau$  (in which a cluster passes through the resonator) are small. The rate of relaxation $\gamma$ of the field  is low, which means a high-Q resonator. The coupling constant  $g$ of the field mode interacting with internal degrees of freedom of the atom is sufficiently large. Trajectories are assumed to be quasiclassical. In more exact terms these assumptions  can be recapitulated as

\begin{equation}
R\tau \ll 1,\,\,\tau \gamma \ll 1,\, g\tau \geq 1\label{1}
\end{equation}

  The micromaser operating on a periodic sequence of $N$-atomic clusters which are created by laser pulses in the gas of unexcited atoms is introduced in \cite{7}. It is assumed that the size of the cluster is much less than the wavelength of the microwave radiation. Effects of finite cluster size \cite{8} are comparable in magnitude with effects of inhomogeneous field at the edges of resonator. The latter observation was reported in Ref. \cite{9}.
In this Letter we study the one-cluster extension of the basic
model. This means that we assume a point-like structure of
$N$-atomic cluster (i.e. the finite size effects are not taken
into account) as well as the fulfillment of the conditions
Eq.(\ref{1}). This formulation generalizes greatly the basic model
while leaves intact the simplifying assumption that process of
interaction of the cluster and the field (within the time interval
$\tau $) and the process of the field relaxation to the
thermodynamic equilibrium (time interval $T\sim 1/R$) are
separated in time. This latter assumption allows to factorize the
evolution operator  of the RDM (see Eq. (\ref{2a}) below) and
 greatly simplifies the analysis of the properties of evolution
operator and the dynamics of RDM.

One-cluster model of the micromaser assumes that the $N$-particle Tavis-Cummings Hamiltonian~\cite{10}
\begin{equation}
H=H_0+V=\omega\left (a^\dag a + S_3+\frac{N}{2}\right)+2g\left(a^\dag S_- +a S_+\right)\label{2}
\end{equation}
 is applicable.  Here $\omega$ is the frequency of the quantum transition which is in exact resonance with
the field mode. The collective spin variables $S_3,S_\pm$ are the generators of the $su(2)$ algebra, while
$a^\dag , a$ are the creation and annihilation operator of the field mode, $\hbar=1$.

The operation of the micromaser for each cluster is divided into two time intervals: the interaction time $\tau $ and the relaxation time $T$. This means that the vector of the main diagonal $\rho^{(l)}$ of RDM
 satisfies the  following equation

\begin{eqnarray}
\rho^{(l+1)}&&=S(N)\rho^{(l)}=Q(N_{ex})Sp_{at}\left( e^{-iH\tau}\rho_{at}\otimes \rho^{(l)}e^{iH\tau} \right)\nonumber\\
&&= Q(N_{ex})W(\tau ) \rho^{(l)}.\label{2a}
\end{eqnarray}

This difference equation connects the main diagonals of RDM taken at the instants when the $l$-th and $(l+1)$-th clusters enter the cavity. This allows to understand the number of passing clusters $l$ as  a discrete "time variable".
Here $N_{ex} =R/\gamma$, and $Q(N_{ex})$ is the evolution operator of RDM at the relaxation stage, i.e. in the
 empty resonator~\cite{5}. The operator $W(\tau)$ describes the evolution of RDM at the stage of interaction of $(l+1)$-th cluster with the field, $\rho_{at}$ is the density matrix of $N$-atomic
cluster before it enters the resonator. The operation $Sp_{at}$ means the trace with respect to atomic
variables.

In our work we consider clusters of fully excited atoms, while the field is initially prepared in the state of the thermal equilibrium with the mean number of photons $n_b=0.1$.
In our forthcoming publication we will rigorously show that if there is an additive with respect to atoms and field integral of motion $[H,H_0]=0$ and for unpolarized initial state of the cluster, then the dynamics of RDM is {\em diagonally invariant}. This important property of the evolution operator $W(\tau)$ means that  each (sub)diagonal of RDM in the Fock basis evolves independently of other elements of RD matrix. In this Letter we concentrate on the dynamics of the main diagonal of RDM, i.e. on the number of photons probability distribution function.
In the space of vectors with components $ \rho^{(l)}_n $, the evolution operators $Q(N_{ex})$  can be represented as
the following matrix

\begin{equation}
Q(N_{ex})=\left(1+\frac{L}{N_{ex}}\right)^{-1},
\end{equation}

where the operator $L$ in the matrix form reads

\begin{equation}
L_{nm}=[-(n_b+1)n_b(n+1)]\delta_{nm}-(n_b+1)(n+1)\delta_{n+1,m}+n_b n\delta_{n-1,m}.
\end{equation}
The matrix of the evolution operator $W(\tau)$ in the Fock basis is low-triangular. In the present Letter we analyze this matrix by numerical methods.
The property of diagonal invariance  simplifies greatly the analysis of RDM dynamics.

The vector of the main diagonal of RDM satisfies the following difference equation

\begin{equation}
   \rho^{(l+1)}_n- \rho^{(l)}_n=  J^{(l+1)}_n- J^{(l)}_n.   \label{3}
\end{equation}

Here ${J}^{(l)}$ is the vector of the probability flux for the $l$-th passage.
The components of this vector are
\begin{equation}
J_n^{(l)}=-\sum_{n'=0}^{n-1}\sum_{n''=n} S(N)_{n'n''}\rho_{n''}^{(l)}+ \sum_{n'=n} \sum_{n''=0}^{n-1} S(N)_{n'n''} \rho_{n''}^{(l)}\label{4}
\end{equation}

This vector  determines the rate of change (after one passage) of the sum of probabilities of the photon numbers in the interval of Fock numbers between $n=n_0$ and $n=n_1$ .  This rate is equal to the difference of fluxes through the chosen boundary values, viz

\begin{equation}
   \sum_{n=n_0}^{n_1}\left( \rho^{(l+1)}_n -\rho^{(l)} _n  \right)  =J^{(l)}_{n_0}-J^{(l)}_{n_1+1} \label{5}
\end{equation}

The dependencies of the eigenvalues $W_n$ on the number of photons
are given in Figure 1 for the number of atoms in the cluster
N=1,5,10. The interaction time is chosen as $g\tau =1.355$ . The
eigenvalues $W_n$ are positive and do not exceed 1. Their mutual
positions are defined by the parameter $\tau$  and the number of
atoms $N$. For the one-atom micromaser the so-called trapped
states are known. These are the Fock states for the number of
photons $n$ corresponding to the eigenvalue $W_n=1$ of the matrix
$W(\tau )$ . This number of photons fulfills the trapping
condition
\begin{equation}
 \sqrt{n+1}=\frac{\pi \chi }{g\tau }\label{6}
\end{equation}
where  $\chi $ is an integer number. The trapped states do not
decay in the absence of relaxation, and thus determine the
dynamics of $\rho^{(l)}$  for large $l$. The recent experimental
realization of the trapped states was reported in~\cite{wal}. The
Figure 1 shows that in the multi-atomic case there are no trapped
states. There are however a few eigenvalues which are close to 1.
The corresponding eigenvectors in the space of the number of
photons are localized around the numbers $n $ for which
$W_n\approx 1$. Such long-living vectors is natural to call {\em
quasitrapped states}.

 The Figure 2 shows the spectrum $S(N)$ in ascending order. It is interesting to
 notice that the eigenvalues of the evolution operator  tend to group around zero when the number of atoms in the cluster increases. In the hierarchy of dynamical processes in the micromaser the small eigenvalues  are responsible for the rapid phase of the dynamics (with respect to the discrete time $l$ ). The quasitrapped states corresponding to the eigenvalues in the interval
$[ 0.9,1)$ are in turn responsible for the slow phase of dynamics.
Probabilities of the states with corresponding photon numbers at
certain stages of the field formation can be rather high. In
Figure 2 we compare the spectrum of the evolution operator $S(N)$
for the cases with ($N_{ex}=20$) and without relaxation. The
Figure 2 shows in particular that for bigger $N$ the spectrum of
$S(N)$ is more stable towards the influence of relaxation. The
relation Eq.(\ref{2a}) describes the transition of the diagonal
elements of RDM to a stationary state. This transition process is
determined by the pumping of the cavity field by passing clusters
as well as by the relaxation of the field. The recent literature
discusses mostly ~\cite{3,5}  the properties of the stationary
state, which can be achieved when a large number of clusters has
gone through the cavity. This case corresponds to the asymptotic
limit $\l \to\infty$ . In this work we concentrate on the
properties of the transition process which, due to the existence
of the quasitrapped states, are very interesting. The field rather
rapidly "forgets" its initial state of the thermal equilibrium.
The dynamics of the population of the Fock states shows instead
the formation of long-living (with respect to the "time" $l$)
quasi-equilibrium distributions. This is illustrated by the
properties of spectrum of the evolution operator given in Figures
1,2. The existence of the eigenvalues close to 1 indicates
considerable probabilities of the Fock states with photon numbers
in the vicinities of the maxima. The small eigenvalues correspond
to the sharp depletion of the corresponding Fock states. The
Figure 1 shows that in the chosen interval of Fock numbers, $0\leq
n\leq 60$, and for $g\tau=1.355$ there are three domains capturing
considerable probabilities. These domains, which are natural to
call the domains of {\em quasitrapping} are localized in the
vicinities of the Fock numbers $n=5,18,40$ and contain almost all
the probabilities. This means that they are getting populated at
different "moments" of "time" $l$  in relays: the next domain
cannot get populated until the previous one is depleted. This
relay of populations is illustrated in Figures 3,4,5. The Figure 3
show for $ N=10$ how the sums of probabilities of the Fock states
change with $l$ in the second ($14\leq n\leq 24$) and the third
($39\leq n\leq 49$) domains of quasitrapping. The rates of
probability change are calculated through  Eq.~(\ref{5}) i.e. as
the flux differences  through the boundaries of the chosen
domains. The Figure 3 allows to identify the following stages of
the $l$-dynamics: a period of accumulation of the probability
which corresponds to the positive values of the probability rate
as well as an extended in time ($l$) period of the negative
probability rates. The lasting nature of the latter period
indicates that the life-time of the quasitrapped states is
considerable. The rate of decay of the second quasitrapped state
($n\approx18$) is approximately the rate of accumulation in the
third state ($n\approx 40$). This means that through a passage of
a cluster the probability is almost fully relayed from the second
quasitrapped state to the third. Since the dynamics of the decay
of the second quasitrapped state is slow, so is the dynamics of
the accumulation in the third state. In Figures 4 and 5 are given
dependencies on $l$ of total populations  curves of the domains of
quasitrapping. It is evident from Figures 4 and 5 that the sum of
populations of two subsequent domains of quasitrapping is close to
1. This again manifests the full accumulation of the probabilities
in the domains of quasitrapping as well as the relay of
probabilities indicated above. The Figure 6 shows the
distributions of the diagonal elements of the RDM taken at the $l$
-moments of maximal probabilities of the Fock states in the
corresponding domains of quasitrapping. As follows from Figure 6
it is possible to govern the vector of photon number distribution.
 This can be achieved by the variation of the number of atoms passed through the resonator.
 It is possible in particular to create states close to the Fock states localized at certain photon numbers.  We can
 also report that a domain of the localization changes smoothly in accord with variations of parameters $\tau$, $N$ and $N_{ex}$ .
The dependence of the stationary field on these parameters for $N=1$ was discussed
in Refs. ~\cite{3,5,6}.
The possibility to engineer quantum states is actively studied in the recent literature~\cite{12}.

\section*{Conclusions and discussion}
The main result of our work is the discovered possibility to
purposefully create in the cavity quasistable states close to Fock
states. We analyzed the dynamics of the micromaser  pumped by
$N$-atomic clusters ~\cite{7}. Our approach generalizes the basic
model of the one atom micromaser ~\cite{3} and can be
experimentally realized. We assumed the point-like nature of
$N$-atomic clusters. This assumption  can easily be realized in
practice when clusters are created in a gas flow by focused laser
pulses in the light range. In this case the width of the beam is
of order of few microns while the size of the cavity can be of
order of few millimeters. In our work we have pointed out the
conditions when the time evolution of a (sub)diagonal of the
reduced density matrix is independent of the other elements of the
density matrix. We have investigated the properties of the
spectrum of the evolution operator (see Figures 1,2) and
 discussed their connection to the properties of the RDM dynamics.  We have discussed the
 hierarchy of the time scales of the micromaser dynamics and have shown that the sectors of the
  spectrum around zero are responsible for rapid processes while the sectors close to 1
  correspond to quasi-equilibrium. For the first time in the existing literature we have
  introduced an important notion of the {\em quasitrapped states}. The Figure 6 shows that
  these states are close to the Fock states. The domains in the Fock space corresponding to
  quasitrapping are rather narrow, their locations change smoothly with
  variations of the number of atoms in the cluster. This means that the overall picture of
  the dynamics is stable with respect to small variations of the number of atoms in a cluster.
  In our future work we plan to investigate this phenomenon in a greater detail as well as to
  study how the properties of the quasitrapped states depend on the choice of the initial
  density matrix of the $N$-atomic cluster.

\pagebreak[4]
\section*{Figure captions}

Figure 1. The spectrum of $W(0,\tau)$  for $N=1,5,10$  and
$g\tau=1.355$ . \vskip 1cm

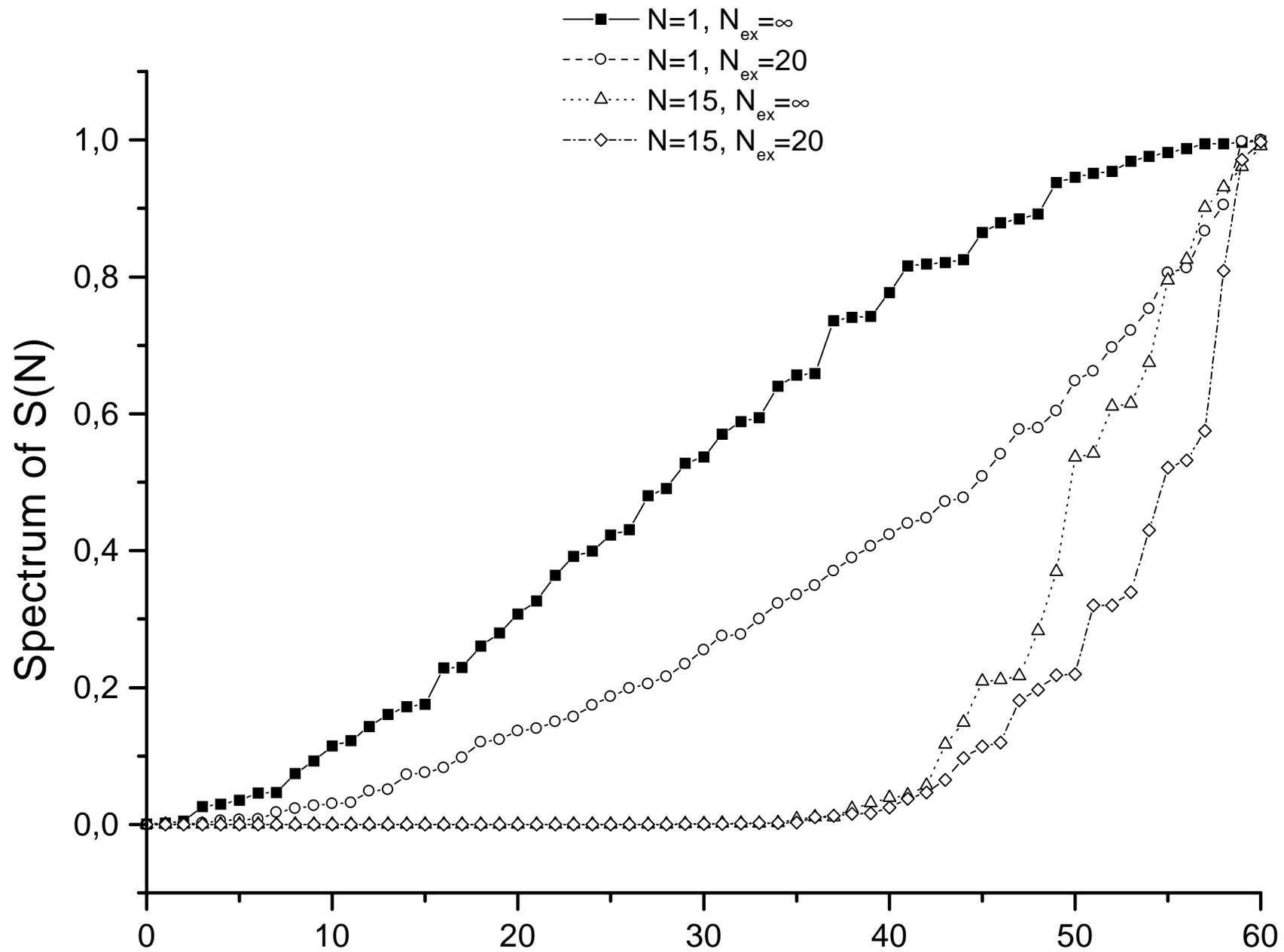
Figure 2. The spectrum of the evolution operator $S(N)$  in
ascending order for $N=1,15$, $N_{ex}=20,\infty$, and
$g\tau=1.355$ . \vskip 1cm

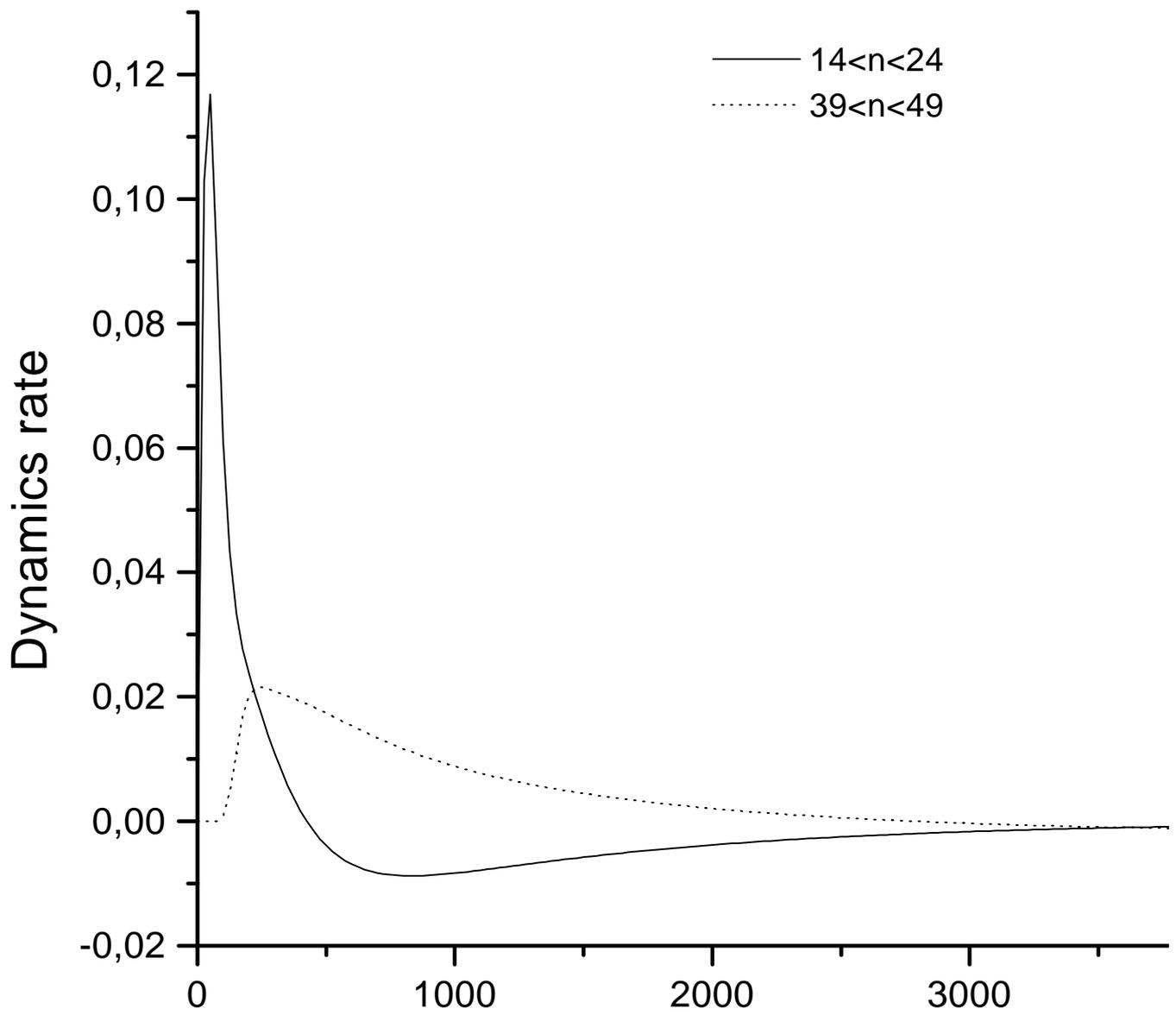
Figure 3. The rates of change of integral probabilities of the
Fock states in the second $14\leq n\leq 24$  and the third $39\leq
n\leq 49$  quasitrapping domains for $N=10$ .

\vskip 1cm

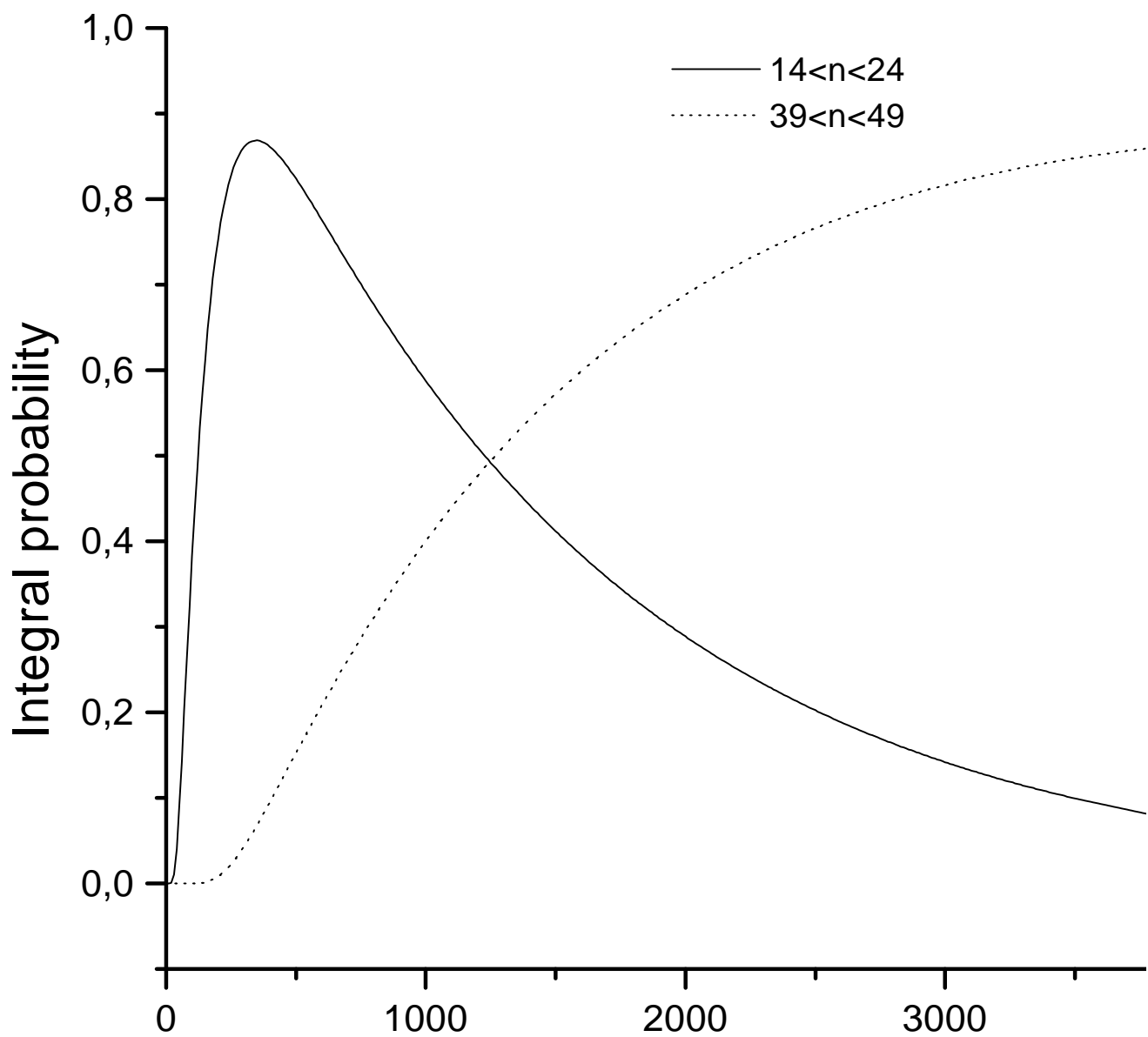
Figure 4.
Integral probabilities of the Fock states in the second
$14\leq n\leq 24$ and the third $39\leq n\leq 49$ quasitrapping
domains for $N=1$ .

\vskip 1cm
Figure 5. Integral probabilities of the Fock states in
the second $14\leq n\leq 24$ and the third $39\leq n\leq 49$
quasitrapping domains for $N=10$ .

\vskip 1cm

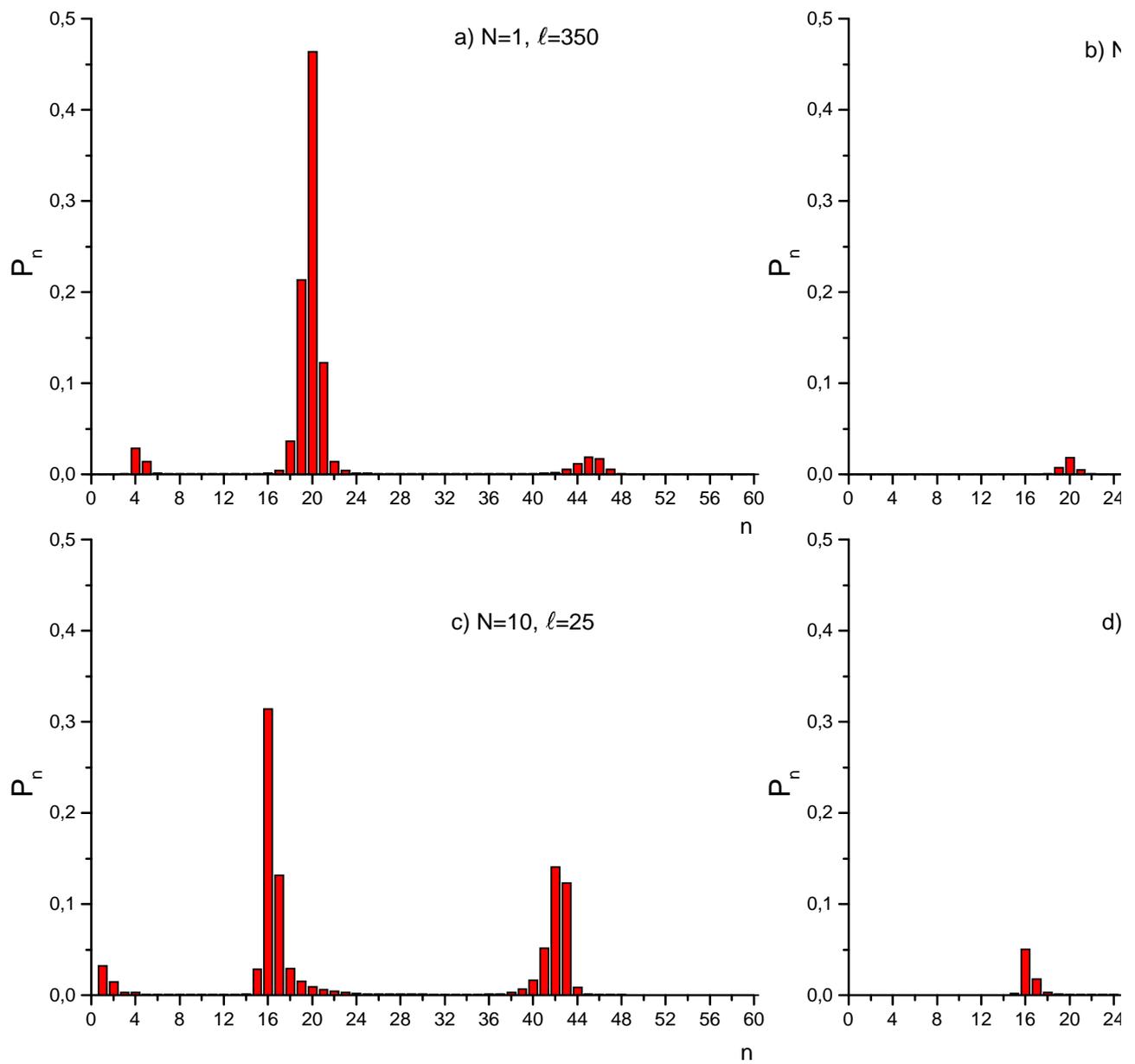
Figure 6. The photon number distributions of the diagonal elements
of RDM at the $l$ -moments of maximal probabilities of the Fock
states in the second $14\leq n\leq 24$ and the third $39\leq n\leq
49$ domains of quasitrapping.


\begin{thebibliography}{99}

\bibitem{1} S. Haroche, D. Kleppner, Phys. Today, {\bf 42}(1), 24(1989).
 \bibitem{2} D. Meschede, H. Walther, G. Muller, Phys. Rev. Lett. {\bf 54}, 551(1985); G. Rempe, M. Scully,   H. Walther, Physica Scripta, {\bf34}, 5 (1991); G.M. Brune, J. Raimond, P. Goy, L. Davidovich, S. Haroche, Phys. Rev. Lett. {\bf59}, 1899 (1987).
\bibitem{3} P. Filipowicz, L. Javanainen, P. Meystre, Phys. Rev. A, {\bf34}, 3077 (1986).
 \bibitem{4} P. Meystre, M. Sargent III. Elements of Quantum Optica, Springer-Verlag, Berlin, 1990.
 \bibitem{5} P. Elmfors, B. Lautrup, B. Shagerstam, Phys. Rev. {\bf54}, 5171 (1996).
 \bibitem{6} P. Meystre, G. Rempe, H. Walther, Opt. Lett. {\bf13}, 1078 (1988).
 \bibitem{7}  G. D'Ariano, N. Sterpi, A. Zucchetti, Phys. Rev. Lett. {\bf74}, 900 (1995).
\bibitem{8}  M. Orszag, R. Ramirez, J. Retamal, C. Saavedra, Phys. Rev. A {\bf49}, 2933 (1994); L. Ladron, M. Orszag, R. Ramirez, Phys. Rev. A {\bf55}, 2471 (1997); M. Kolobov, F. Haake, Phys. Rev. A {\bf 55}, 3033 (1997).
  \bibitem{9} C. Yang, K. An, Phys. Rev. A {\bf55}, 4492 (1997); F. Gasagrande, A. Lulli, S. Ulrega,
Phys. Rev. A {\bf60}, 1582 (1999).
\bibitem{10} M. Tavis, E. Cummings, Phys. Rev. {\bf170}, 379 (1968);
  M. Sculle, G. Meyer, H. Walther, Phys. Rev. Lett. {\bf76}, 4144 (1996);  A. Rybin, G. Kastelewicz, J. Timonen, N. Bogoliubov, J. Phys. A: Math. And Gen. {\bf 31}, 4705 (1998).
\bibitem{wal} M. Weidinger, B.T.H. Varcoe, R. Heerlein, and H.
Walther, Phys. Rev. Lett. {\bf 82}, 3795 (1999).
Phys. Rev. Lett. {\bf82}, 3795 (1999).
\bibitem{12} K. Vogel, V. Akulin, and W. Schleich, Phys. Rev. Lett. {\bf 71}, 1816 (1993); Shi-Biao Zeng, Guang-Gan Guo, Phys. Lett. A, {\bf 244}, 512 (1998); A. Kozhekin, G. Kurizki, and B. Sherman, Phys. Rev. A, {\bf 54}, 3535 (1996).
\end{thebibliography}
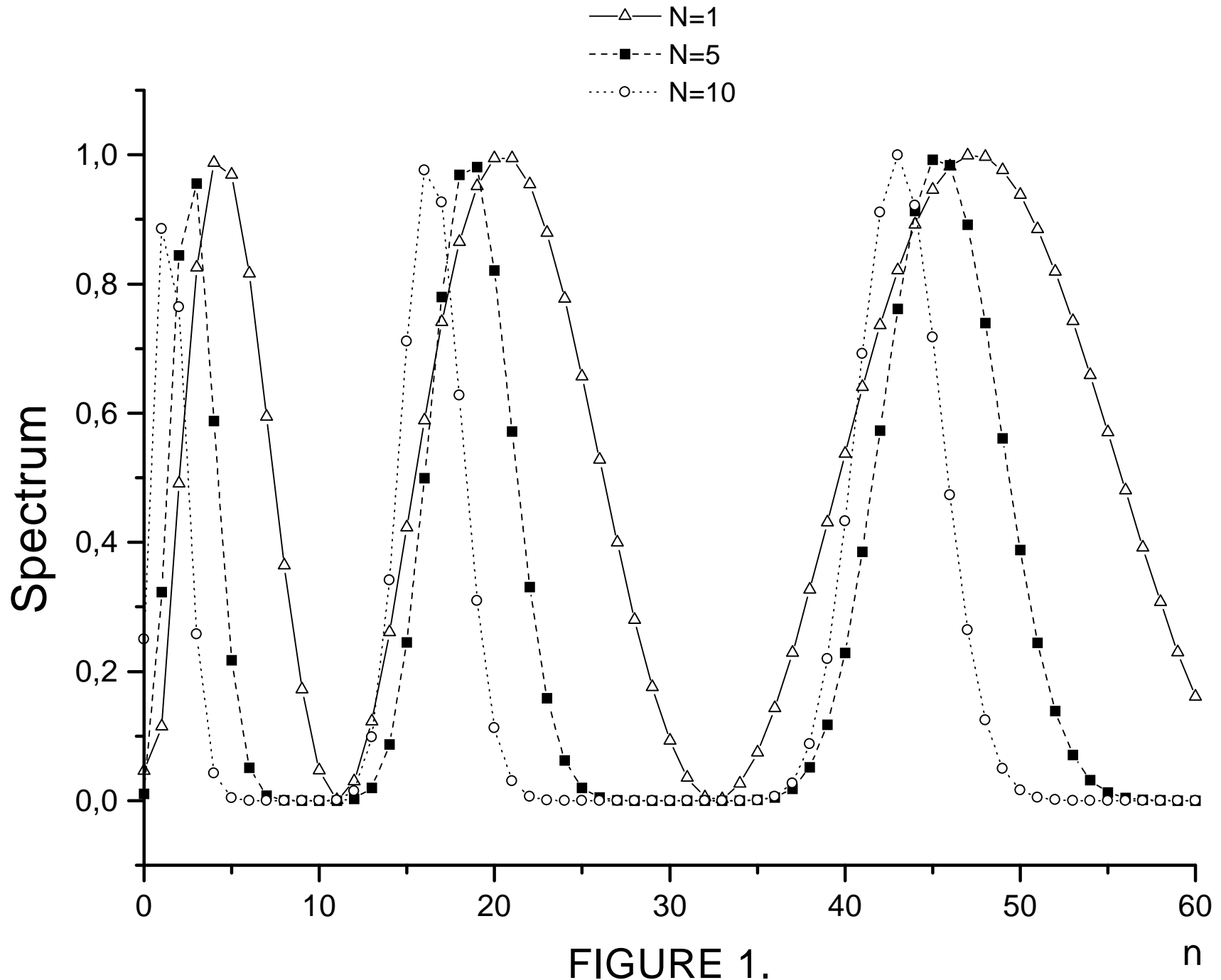
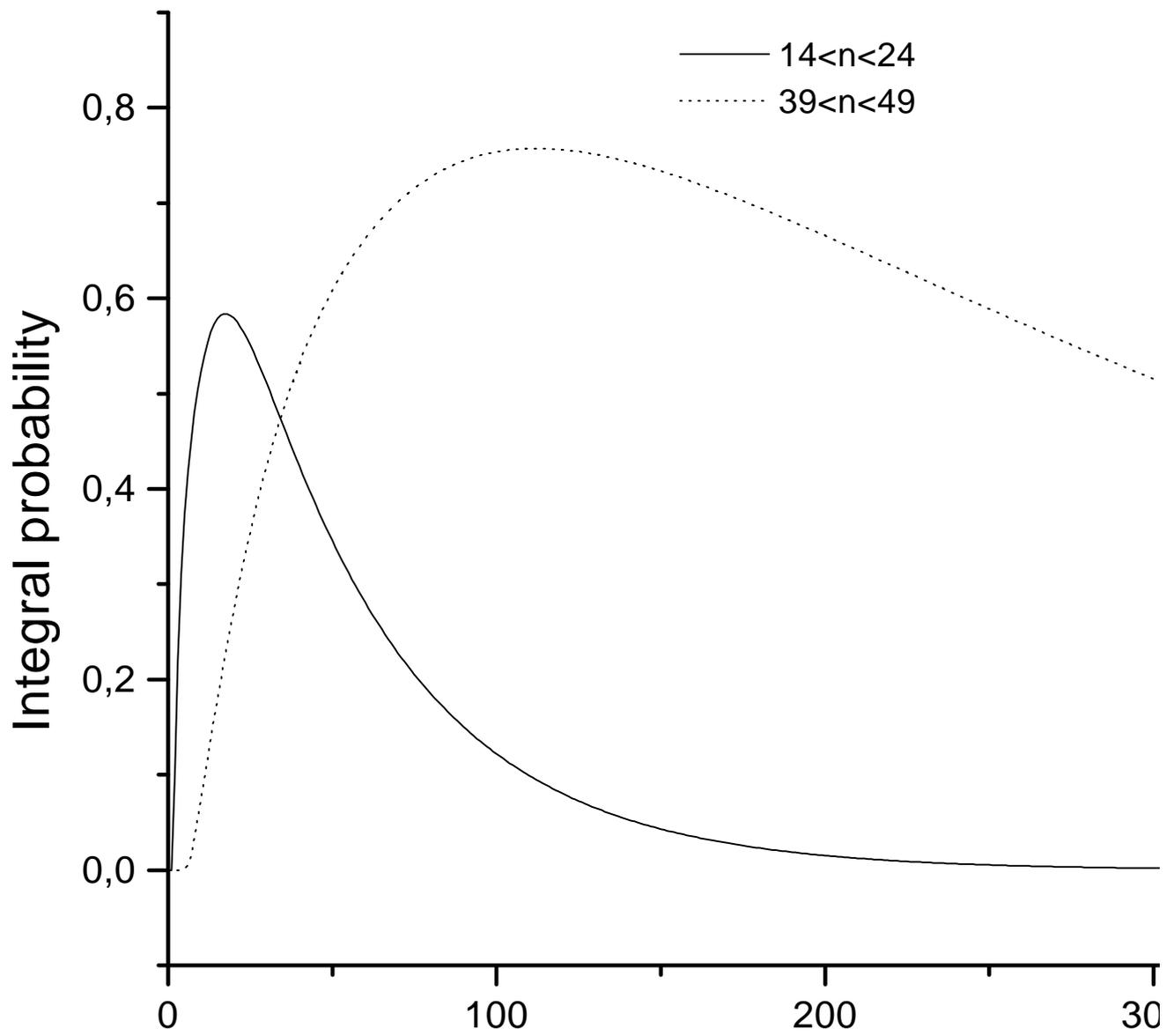
\end{document}